# Implementation of Correlation and Regression Models for Health Insurance Fraud in Covid-19 Environment using Actuarial and Data Science Techniques

Rohan Yashraj Gupta, Satya Sai Mudigonda, Pallav Kumar Baruah, Phani Krishna Kandala

*Abstract: Fraud acts as a major deterrent to a company's growth if uncontrolled. It challenges the fundamental value of "Trust" in the Insurance business. COVID-19 brought additional challenges of increased potential fraud to health insurance business. This work describes implementation of existing and enhanced fraud detection methods in the pre-COVID-19 and COVID-19 environments. For this purpose, we have developed an innovative enhanced fraud detection framework using actuarial and data science techniques. Triggers specific to COVID-19 are identified in addition to the existing triggers. We have also explored the relationship between insurance fraud and COVID-19. To determine this we calculated Pearson correlation coefficient and fitted logarithmic regression model between fraud in health insurance and COVID-19 cases. This work uses two datasets: health insurance dataset and Kaggle dataset on COVID-19 cases for the same select geographical location in India. Our experimental results shows Pearson correlation coefficient of 0.86, which implies that the month on month rate of fraudulent cases is highly correlated with month on month rate of COVID-19 cases. The logarithmic regression performed on the data gave the r-squared value of 0.91 which indicates that the model is a good fit. This work aims to provide much needed tools and techniques for health insurance business to counter the fraud.*

## I. INTRODUCTION

Fraud is malpractice, an act of using a dishonest method that is done to gain some financial benefits, which are not otherwise entitled [1]. It is a major problem in many financial and non-financial sectors which includes providing wrong (misleading) information, raising a false claim, etc [2]–[4]. Today, economies all over the world are plagued with fraud that is affecting various aspects of organizations ranging from financial performance to organizational morale [5].

Insurance fraud is not new to this world. This came into existence ever since insurance business took the form of a commercial enterprise. According, to report by "Coalition Against Insurance Fraud" around 80 billion dollars of fraud is committed each year across all lines of insurance business [6].

According, to an article by Insurance Business Asia, the Indian non-life market has seen a market growth of 14.5% [7], this growth has also generated a humongous amount of data. Unfortunately, a majority of companies have legacy systems that do not capture sufficient details to identify fraud. In the process, companies identify very few cases of fraud and often it is years later that these cases come into the light[8], [9]. Some companies, on the other hand, have leveraged this data to improve their fraud management mechanisms, thereby gaining a competitive advantage over their peers[10]–[12].

Today, due to COVID-19 the current scenario is different compared to what it was earlier. There are reports which state that insurers can expect higher fraudulent claims in insurance. This could bring additional challenges for insurers as the existing models may not be sufficient to detect or prevent fraud which will arise out of this situation.

The various effect of COVID-19 on insurance business is listed below:

- Changes to terms and conditions of the coverage which triggers loss
- Requesting for coverage extension
- Interpretation of the coverage
- One-off payments such as ex-gratia – setting a threshold
- Actuarial Control Cycle - general commercial and environmental changes
  - economic cycle - no business activity → more litigation → more crime → fraudulent claims
  - underwriting cycle - changing from soft to hard market i.e. lesser competition, frequency of claims is higher most claims do not meet standard criteria

Solvency - more liabilities so company alters balance to show solvency. More loss → less profit → not solvent → window dressing[3], [10], [13]–[18]

* Correspondence Author

**Rohan Yashraj Gupta***, Department of Mathematics and Computer Science, Sri Sathya Sai Institute of Higher Learning, Puttaparthi, India. Email: rohanyashrajgupta@sssihl.edu.in

**Satya Sai Mudigonda**, Department of Mathematics and Computer Science, Sri Sathya Sai Institute of Higher Learning, Puttaparthi, India. Email: satyasaibabamudigonda@sssihl.edu.in

**Pallav Kumar Baruah**, Department of Mathematics and Computer Science, Sri Sathya Sai Institute of Higher Learning, Puttaparthi, India. Email: pkbaruah@sssihl.edu.in

**Phani Krishna Kandala**, Department of Mathematics and Computer Science, Sri Sathya Sai Institute of Higher Learning, Puttaparthi, India. Email: kandala.phanikrishna@gmail.com

# Implementation of Correlation and Regression Models for Health Insurance Fraud in Covid-19 Environment using Actuarial and Data Science Techniques

This work focuses on applying actuarial and data science techniques to identify fraudulent cases in COVID-19 environment and establishing its correlation with COVID-19 cases [15], [16], [19]–[23]. There are IX sections in this paper. Section II explains the motivation for this work. Section III provides details about the implementation of existing framework pre-COVID-19. Section IV explains the implementation of the enhanced framework to identify fraud during COVID-19 environment. Section V discusses correlation model for health insurance fraud and COVID-19 cases. Section VI discusses regression models for health insurance fraud and COVID-19 cases. Section VII describes the interpretation of the results. Section VIII states the conclusions based on the work done. Section IX provides a preview of the future work that is being undertaken in this area.

## II. LITERATURE REVIEW AND MOTIVATION

Fraud is one of the most expensive crime and the effect of it is felt directly by customers and various other stakeholders. Due to fraudulent activities, companies face a huge amount of losses [24][25]. These losses invariably result in an increased premium for future customers. Fraudsters employ various types of techniques, strategies and tools to commit fraud. Some of the most common types of fraud include Health care fraud, Debit and Credit card fraud, Identity Theft, Health Insurance fraud, etc.[6][11] Many stakeholders get affected either directly or indirectly because of fraud. A direct impact that is faced by the insured is the increase in premium; this could lead to a knock-on effect for the insurance companies by a loss in business [26]. It thus becomes very important for a corporate entity to have an effective fraud management process to ensure a healthy financial future. A key element that is required to identify and combat fraud is access to data and systems. If one has access to the right set of data fields and has systems equipped with sufficient controls, fraud can be managed better [3], [17], [18], [23], [27]. Fraud has become a cause of anxiety for many organizations. In many companies, fraud is, in most cases, identified only after it has occurred. Ideally, we should be able to identify fraud before the damage is done (i.e. identifying proactively). Fraud detection and as a consequence prevention will help save organizations large portion of their earnings. This will also increase the confidence of the organization. A strong fraud prevention system will increase the confidence of all the stakeholders including the investor and customer towards the company. However, existing fraud detection methods may be insufficient due to challenges posed by COVID-19 [28]. Based on the most relevant studies done in the area of insurance fraud detection, we have identified various methodologies which are currently being used. We analysed this further and arrived at fraud detection and prevention framework incorporating both actuarial and data science techniques. Now, this gets enhanced to meet additional challenges posed by COVID-19 [21], [29], [30]. The concept is depicted in Figure 1

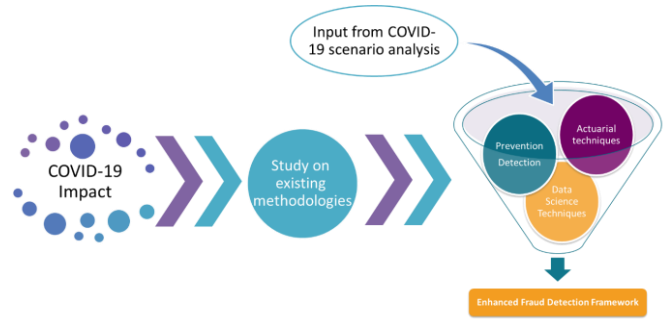

**Figure 1 – Fraud detection concept**

In the past, we had seen how the financial crisis of 2008 which lead to recession triggered many individuals to commit fraud. Association of British Insurers (ABI) reported 107,000 claims as false amounting to a total of £730m during that period. Looking at the past, insurers can expect similar situation arising out of COVID-19 which will trigger more fraudulent claims. The recent study done by Google on search query showed that there was 125% increase in the number of users looking for information on how to start a fire, in the last week of March [31] [19]. This indicates that people are finding ways of collecting money through insurance by committing self-staged accidents.

In our previous work we have developed a framework for fraud detection [29]. This has been enhanced considering the new challenges and risks that COVID-19 poses. COVID-19 specific triggers are considered and presented in section IV(A).

## III. IMPLEMENTATION OF EXISTING FRAMEWORK PRE-COVID-19

The framework for fraud detection in pre- COVID-19 environment is given in Figure 2 [29]:

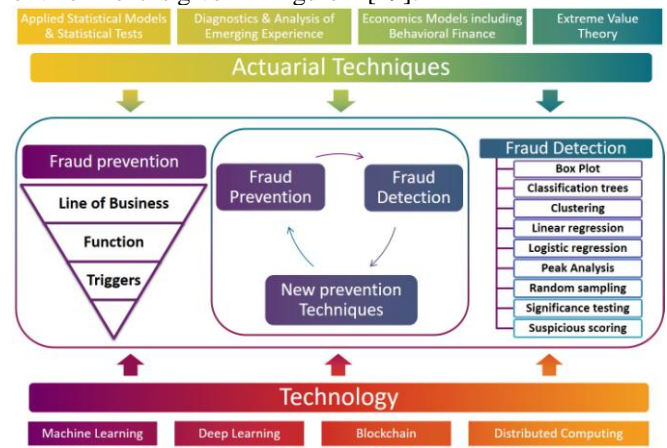

**Figure 2 – Existing fraud detection framework**

The implementation of the framework in the pre-COVID-19 environment using triggers for fraud prevention and data science techniques for fraud detection is shown below for the health insurance dataset [21], [29].

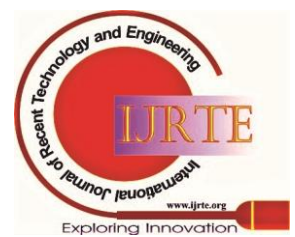

## Data Description

The health dataset used has the data from August, 2019 to August, 2020. The total number of records are 3,76,402 with a total identified fraud cases 29,038. The data has a total of 26 features, details of which is given in the Table 1

**Table 1 – Data Description**

| Feature | Description |
|---------|-------------|
| Policy Number | Unique policy identification number |
| Insured Id | Unique ID given to insured |
| Claim Identification Number | Unique claim identification number |
| Benefit Type | Type of benefit (medical / surgical) |
| Claim Status | Current status of the claim |
| Treatment Start Date | Treatment Start Date |
| Treatment End Date | Treatment End Date |
| Claim Settlement Date | Date at which the claims were settled |
| Claim Reported Date | Date at which the claims were reported |
| Claim Billed Amount | Claims amount billed |
| Approved / Allowed Amount | Approved / Allowed Amount |
| Claim Paid Amount | Claims amount paid |
| Medical Service Provider ID | Unique ID given to hospitals |
| Medical Service Provider Name | Name of the hospital |
| No of Days Stayed | Days stayed in the hospital |
| Primary Diagnosis Code | Unique code given to diagnosis |
| Primary Diagnosis Name | Name of the diagnosis |
| Primary Procedure Code | Unique code given to procedure |
| Primary Procedure Name | Name of the procedure |
| Net Amt | Net amount paid to the insured |
| Claim Paid Date | Date of payment of the claim |
| Surgery Date | Date of surgery |
| Discharge Date | Date of discharge |
| Claim Raised Date | Date when the claim was raised |
| Hospital District | District where the hospital is located |
| Fraud Status | Status of the claims as fraud / not-fraud |

### A. Implementation for fraud prevention using triggers

A mechanism for early detection of the features of a fraudulent activity will give the company an edge to develop effective fraud prevention and management system. Such features are termed as triggers. Identification of these triggers in various activities within the company would help in taking required measures for prevention. A tool for detecting triggers on real time basis is going to be very useful for the organization.

Some of the triggers identified pre-COVID-19 environment are as follows:

- Abnormally long time off for a given injury
- Multiple medical provider/ opinions
- Applicant fails to sign and put the date in the application
- Submitting fake test reports
- Visible tampering of document
- Hand written bills
- High value bills (very high doctor fee)

It may not be possible to detect fraud in the early stages of the process in the normal work flow of the organization. However, it is possible to build a model for early detection of fraud by using data science techniques.

### B. Implementation for fraud detection using a predictive model

Data from a health insurance industry is considered for fraud detection analysis. Gradient boosting method is used to build a fraud detection model [21]. The dataset was divided into train and test in the ratio of 70:30. To assess the performance of the model the Receiver Operating Characteristic (ROC) curve is drawn. ROC curve is the plot between true positive rate and false positive rate for different threshold. Each point of the graph represents a sensitivity/ specificity pair for a given threshold point. A model is said to perform good if the area under the curve (AUC) is close to 1. For the current model, the AUC value was found to be 0.9243. This means that the model has optimal performance in the given scenario [21].

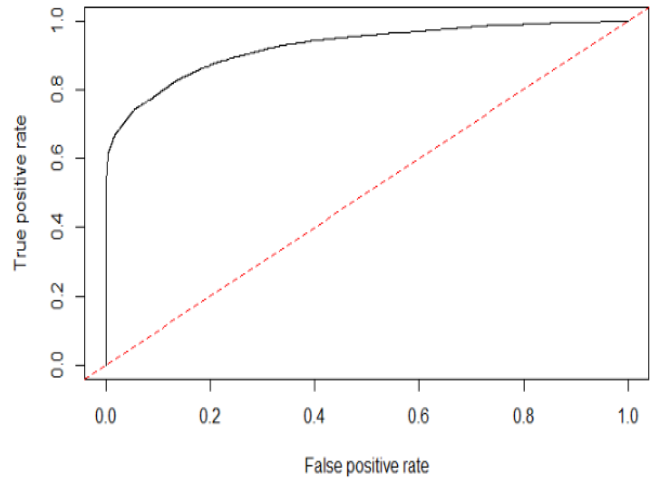

**Figure 3 – ROC Curve of the model**

The other most useful metric used to assess a class imbalance classification problem is F1-score [30]. The F1-score for this model was found to be 0.7191.

However, under COVID-19 scenario the existing model may not perform as expected (shown in the later section). This is because there may be a completely new type of fraud that is occurring. The model may need to be retrained to handle the scenario.

### C. Results of implementation pre-COVID-19

Based on the triggers identified in section III(A) and predictive model in section III(B) a summary of fraudulent cases in the pre-COVID-19 environment is obtained and shown in Table 2

**Table 2 – Fraud data pre-COVID-19**

| Month | Fraud Rate |
|-------|------------|
| Aug-19 | 0.58% |
| Sep-19 | 1.08% |
| Oct-19 | 1.19% |
| Nov-19 | 2.65% |
| Dec-19 | 5.14% |
| Jan-20 | 6.86% |
| Feb-20 | 4.21% |

Where,

$$\text{Fraud rate (month)} = \frac{\text{Total fraud cases (month)}}{\text{Total reported claims (month)}}$$

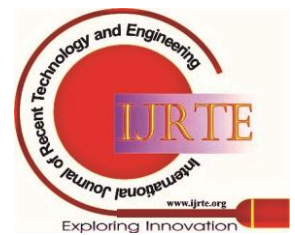

## IV. IMPLEMENTATION OF AN ENHANCED FRAMEWORK FOR FRAUD IN COVID-19 ENVIRONMENT

An appropriate framework is designed based on the newly identified triggers as described in the previous section. This framework is an enhanced and updated version of our previous works [21], [29]

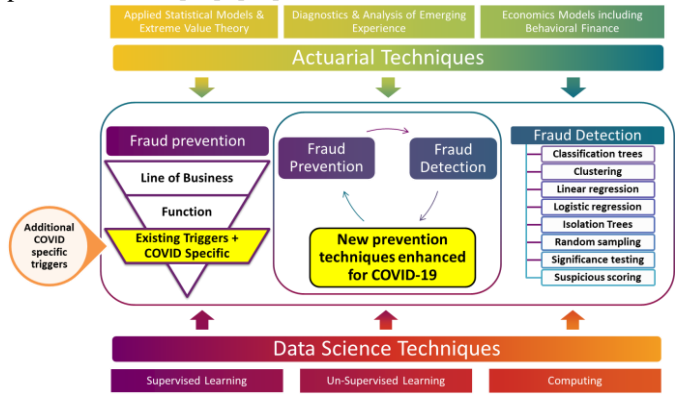

**Figure 4 – Enhanced framework in COVID-19 environment**

The central part of the Framework includes fraud prevention, detection and identification of new prevention techniques supported by both actuarial and data science techniques.

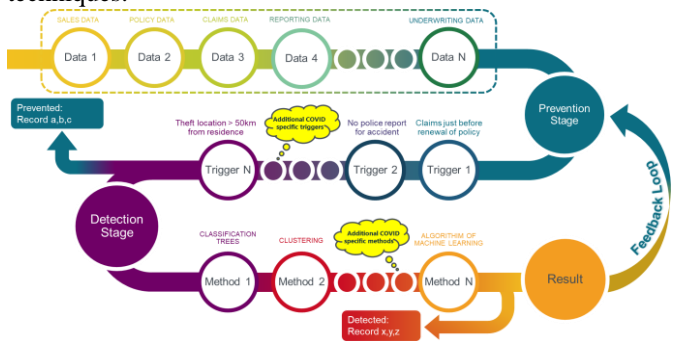

**Figure 5 – Enhanced framework business view**

Figure 5 shows the business view of the enhanced framework. In the trigger-based prevention stage, additional triggers specific to COVID-19 have been considered [29]. The results obtained are analysed further to improve the existing model. There is a feedback mechanism which requires the business experts to exercise judgement and add new triggers to the existing model.

The following sub-sections discuss the implementation of fraud prevention and detection in the COVID-19 environment.

### A. Implementation for fraud prevention using triggers during COVID-19 environment

The first stage of the framework comprises of originating preventive techniques for potential fraud using a trigger-based system. Here, the existing triggers need to be enhanced with COVID-19 specific triggers. Triggers are parameters or a set of parameters, which can help, identify and raise alerts for suspicious activities. These can be managed through an automated system or manually. Identification of triggers in itself is an enormous task and requires a thorough understanding of the business processes. Triggers identified could differ based on the stage in which

the fraud is identified. Some examples include:

- Requesting for coverage extension – more than 30 days
- Requesting for an automatic renewal
- Requesting for one-off payments such as ex-gratia – more than 10 lakhs
- Reduction of minimum and deposit premiums by more than 5% in comparison to earlier years
- Non-tracking of exposure – decrease in policy count by more than 10% in the recent 3 months
- Non-isolation – lack of separation between COVID-19 affected lines of business such as travel from the portfolio

Some of the triggers that were used for identifying fraud in COVID-19 environment are listed below:

**Fraud & Abuse related**

- Package already taken by same insured with a different claim ID
- Specific claim has stayed in rejected status for more than 3 months
- Packages used in COVID-19 environment which looks suspicious
- Highly utilized package were used even more during this time. The excess is identified as fraud.

**Process related**

- Claim is submitted after 15 days of Discharge of the patient, therefore claim has been rejected
- Discharge documents submitted after 15 days of discharge
- Procedure not performed

**Eligibility related**

- Wrong package blocked
- Oral medication not payable
- Sign and stamp of the district hospital in referral slip not provided

### B. Implementation for fraud detection using a predictive model during COVID-19 environment

The important step in fraud detection is the identification of suspicious activities that have a higher probability of being fraudulent. Detecting an insurance fraud and abuse requires an in-depth knowledge of the insurance industry. Many insurance systems have experts who manually review each of the transactions to check for the suspicious ones. There are various methodologies that are being used in insurance fraud detection in recent times, which are as follows:

- Data Science techniques
  - Supervised learning
    - Classification trees
    - Linear regression
    - Logarithmic regression
    - Isolation trees
    - Support vector machine
  - Unsupervised learning
    - Clustering
- Actuarial techniques
  - Random sampling
  - Significance testing
  - Suspicious scoring
  - Applied Statistical Models & Statistical Tests
  - Diagnostics & Analysis of Emerging Experience
  - Economics Models including Behavioural Finance
  - Extreme Value Theory

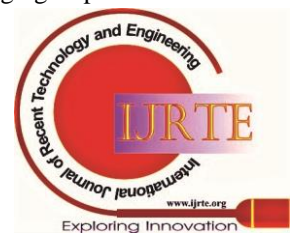

Similar to the experiment done in the pre-COVID-19 secnario[21], the gradient boosting method is run on the health insurance data during COVID-19 environment. Figure 6 shows the receiver operating characteristic plot of the model.

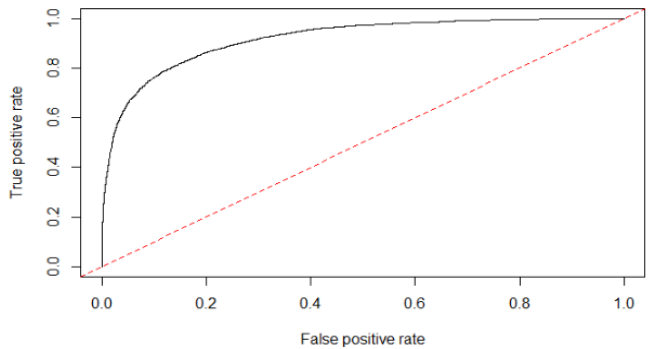

**Figure 6 – ROC curve of the pre-COVID-19 model**

It is observed that the area under the curve of the model was 0.9169, which is lower than what we observed in pre-COVID-19 implementation of the same model. For this implementaion the F1-score was observed to be 0.6486.

Based on the triggers and predictive model a summary of fraudulent cases during COVID-19 environment is given in Table 3 as a percentage of total reported claims

**Table 3 – Fraud data during COVID-19 environment**

| Month | Fraud rate |
|-------|-----------|
| Mar-20 | 6.16% |
| Apr-20 | 6.84% |
| May-20 | 8.51% |
| Jun-20 | 9.89% |
| Jul-20 | 11.89% |
| Aug-20 | 13.96% |

### C. Determining the rate of COVID-19 cases

The COVID-19 dataset used for this purpose is retrieved from Kaggle [32]. This data contains day to day numbers of COVID-19 infected cases, recovered cases and deaths for all the states of India. For this work, we have taken infected cases data from 1$^{st}$ March to 31$^{st}$ August which refers to the same geographical location as of health insurance dataset. Figure 7 shows the plot of the total infected cases of COVID-19 for this period.

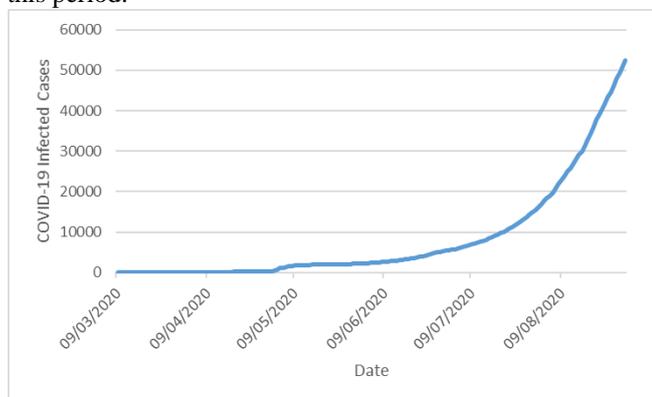

**Figure 7 – Plot of COVID-19 cases**

The month-wise percentage of COVID-19 cases with respect to total population is shown in Table 4

**Table 4 – Month-wise COVID-19 infected cases**

| Month | COVID-19 cases | COVID-19 rate |
|-------|----------------|---------------|
| Mar-20 | 41 | 0.00137% |
| Apr-20 | 316 | 0.01053% |
| May-20 | 1,876 | 0.06253% |
| Jun-20 | 3,185 | 0.10617% |
| Jul-20 | 10,038 | 0.33460% |
| Aug-20 | 37,070 | 1.23567% |

$$\text{COVID-19 rate (month)} = \frac{\text{Total COVID-19 cases (month)}}{\text{Total population (month)}}$$

The estimated total population of the geographical location is 3 million.

## V. CORRELATION MODEL FOR HEALTH INSURANCE FRAUD AND COVID-19 CASES

In this section, we derive the correlation coefficient for health insurance fraud and COVID-19 cases using Pearson correlation coefficient. Table 5 is the consolidation of the data presented in Tables 2, 3 and 4[32]. This depicts month-wise rate of fraud and COVID-19 cases.

**Table 5 - Month-wise data of Fraud and COVID-19**

| Month | Fraud rate | COVID-19 rate |
|-------|-----------|---------------|
| Aug-19 | 0.58% | 0.00000% |
| Sep-19 | 1.08% | 0.00000% |
| Oct-19 | 1.19% | 0.00000% |
| Nov-19 | 2.65% | 0.00000% |
| Dec-19 | 5.14% | 0.00000% |
| Jan-20 | 6.86% | 0.00000% |
| Feb-20 | 4.21% | 0.00003% |
| Mar-20 | 6.16% | 0.00137% |
| Apr-20 | 6.84% | 0.01053% |
| May-20 | 8.51% | 0.06253% |
| Jun-20 | 9.89% | 0.10617% |
| Jul-20 | 11.89% | 0.33460% |
| Aug-20 | 13.96% | 1.23567% |

Since, the COVID-19 cases was negligible for the month of February, we have ignored this and considered data from March to August. The descriptive statistics of the six months data from 1$^{st}$ March to 31$^{st}$ August for both COVID-19 cases and fraud cases are given in Table 6

**Table 6 - Descriptive statistics of the COVID-19 and fraud data**

| COVID-19 rate | | Fraud rate | |
|---------------|---|-----------|---|
| Mean | 0.002918111 | Mean | 0.09543601 |
| Standard Error | 0.001952094 | Standard Error | 0.01224767 |
| Median | 0.0008435 | Median | 0.09201897 |
| Standard Deviation | 0.004781635 | Standard Deviation | 0.030000542 |
| Sample Variance | 2.2864E-05 | Sample Variance | 0.000900033 |
| Range | 0.012343 | Range | 0.078036001 |
| Minimum | 1.36667E-05 | Minimum | 0.061601365 |
| Maximum | 0.012356667 | Maximum | 0.139637366 |
| Count | 6 | Count | 6 |

This gives a range of summary statistics like mean, variance, range, median, etc. which gives a good idea about the behaviour of fraud and COVID datasets.

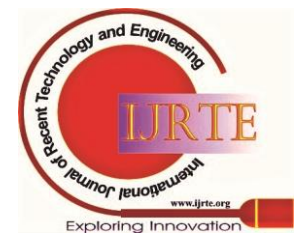

Pearson correlation coefficient is calculated between the COVID-19 cases and fraud cases in health insurance, which is given in Table 7. The formula used to find the Pearson correlation coefficient is as follows:

$$r = \frac{\sum(x_i - x_{average})\,(y_i - y_{average})}{\sqrt{\sum(x_i - x_{average})^2 * \sum(y_i - y_{average})^2}}$$

**Table 7 - Correlation matrix between COVID-19 and fraud cases**

|  | COVID-19 % | Fraud % |
|---|---|---|
| COVID-19 % | 1 |  |
| Fraud % | 0.8626 | 1 |

## VI.  REGRESSION MODELS FOR HEALTH INSURANCE FRAUD AND COVID-19 CASES

Linear and logarithmic regression is fitted considering data shown in Table 5 for the month of March to August.

The regression line thus derived is given as:

Linear regression:

y = 5.4124x + 0.0796

where,

y → Predicted fraud cases %

x → Infected COVID-19 cases %

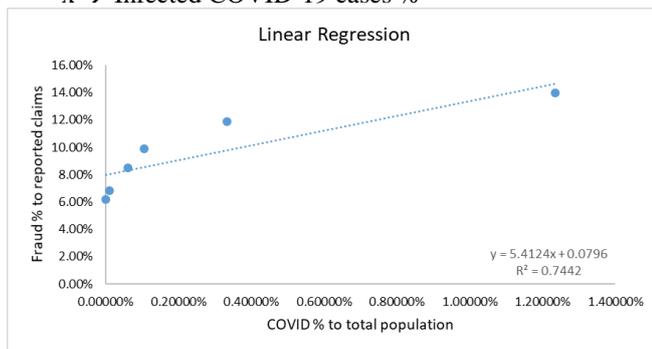

**Figure 8 – Linear regression**

Logarithmic regression:

y = 0.0118ln(x) + 0.1832

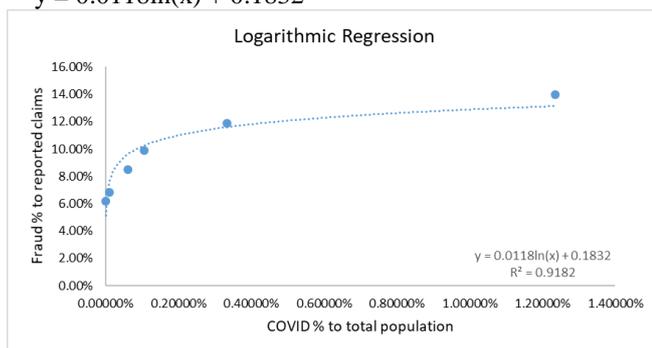

**Figure 9 – Logarithmic regression**

Table 8 shows the summary output of the regression that was performed.

**Table 8 - Summary output of the regression model**

|  | Linear Regression | Logarithmic Regression |
|---|---|---|
| Multiple R | 0.5538 | 0.8431 |
| R Square | 0.7442 | 0.9182 |

Based on the regression line the predicted fraud cases with the corresponding residuals along with the residual plots are given in Table 9 and Figure 10:

**Table 9 – Prediction and residuals for both the regression models**

| Month | COVID-19 % | Linear regression prediction | Logarithmic regression prediction | Linear regression residual | Logarithmic regression residual |
|---|---|---|---|---|---|
| Mar-20 | 0.00137% | 7.97% | 5.10% | -1.81% | 1.06% |
| Apr-20 | 0.01053% | 8.02% | 7.51% | -1.18% | -0.67% |
| May-20 | 0.06253% | 8.30% | 9.61% | 0.21% | -1.10% |
| Jun-20 | 0.10617% | 8.53% | 10.24% | 1.36% | -0.35% |
| Jul-20 | 0.33460% | 9.77% | 11.59% | 2.12% | 0.30% |
| Aug-20 | 1.23567% | 14.65% | 13.14% | -0.68% | 0.83% |

The results from Table 9 show that the residuals obtained from logarithmic regression is much lesser compared to linear regression.

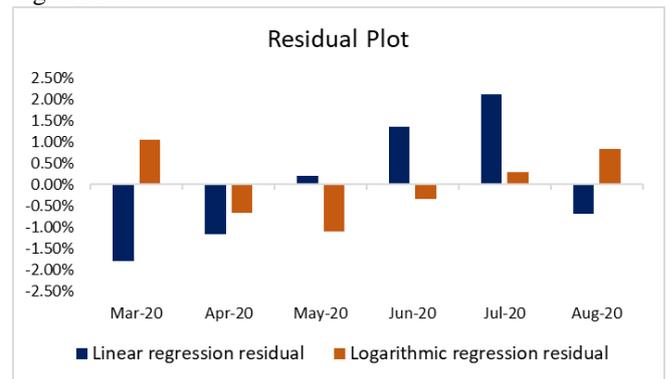

**Figure 10 – Residual plot of the linear and logarithmic model**

Figure 10 shows that the residual plot is randomly distributed for logarithmic regression, which indicates that the logarithmic model is a better fit.

## VII.  INTERPRETATION OF RESULTS

Correlation coefficient is used to determine the relationship between the two variables used. It can take the value from the range -1 to +1. The correlation between COVID-19 and fraud cases is observed to be 0.86. This means that there is a strong positive correlation between the two variables. This implies that with the increase in COVID-19 cases there is a high degree of chance for the increase of fraudulent cases in health insurance business. One of the major reasons for this could be that lockdown during this period has forced many people out of employment. This in turn might have induced the idea in many people of earning money by whatever means available.

Upon performing linear and logarithmic regressions in the dataset, we observed the r-squared value of 0.74 and 0.91 respectively. It is a measure which is used to determine the percentage of variation for a response variable that is explained by the explanatory variable. For the results achieved, it was observed that both the models have a significantly high r-squared value. However, in comparison logarithmic regression is a better fit of the two.

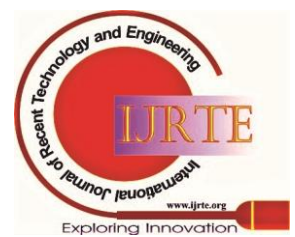

## VIII. CONCLUSION

In this paper, we developed an enhanced fraud detection and prevention framework, which integrates both actuarial and data science techniques. We have shown that there is an increase in fraud rate during COVID-19 scenario. We have identified various triggers to identify fraud specific to the business environment during COVID-19. We have also demonstrated the implementation of fraud detection during COVID-19 environment. We have established that there is a high positive correlation between infected cases and fraud during COVID-19 and also performed regression. We have demonstrated this by using statistical models and machine learning techniques applied on health insurance data. Results indicate that increase in infected COVID-19 cases has a strong positive correlation with a Pearson coefficient of 0.86 to increase in rate of fraudulent cases and that logarithmic regression is a better fit with R-squared value of 0.91.

## IX. FUTURE WORK

Future work includes:
- Implementing enhanced framework for other lines of business
- Quantifying the increase in health insurance fraud during COVID-19 environment
- Providing risk management strategies for increased fraud risk due to COVID-19

## ACKNOWLEDGEMENTS

We dedicate this work to the Revered Founder Chancellor of Sri Sathya Sai Institute of Higher Learning, Bhagawan Sri Sathya Sai Baba.

## REFERENCE

1. R. A. Bauder and T. M. Khoshgoftaar, "Medicare Fraud Detection Using Machine Learning Methods," in *2017 16th IEEE International Conference on Machine Learning and Applications (ICMLA)*, 2017, vol. 2018-Janua, pp. 858–865, doi: 10.1109/ICMLA.2017.00-48.
2. L. V. . M. R. M. Ekin Tahir; Fulton, T. Ekin, R. M. Musal, and L. V. Fulton, "Overpayment models for medical audits: multiple scenarios," *J. Appl. Stat.*, vol. 42, no. 11, pp. 2391–2405, Nov. 2015, doi: 10.1080/02664763.2015.1034659.
3. A. Gangopadhyaya and A. B. Garrett, "Unemployment, Health Insurance, and the COVID-19 Recession," *SSRN Electron. J.*, Apr. 2020, doi: 10.2139/ssrn.3568489.
4. D. Thornton *et al.*, "Predicting Healthcare Fraud in Medicaid: A Multidimensional Data Model and Analysis Techniques for Fraud Detection," *Procedia Technol.*, vol. 9, pp. 1252–1264, 2013, doi: 10.1016/j.protcy.2013.12.140.
5. "Insurance Fraud in Times of Crisis - FRISS." [Online]. Available: https://www.friss.com/blog/insurance-fraud-in-times-of-crisis/. [Accessed: 05-Jun-2020].
6. "Coalition Against Insurance Fraud: Rapid National Response Urged To Head Off Coming Wave of COVID-19 Insurance Scams - InsuranceNewsNet." [Online]. Available: https://insurancenewsnet.com/oarticle/coalition-against-insurance-fraud-rapid-national-response-urged-to-head-off-coming-wave-of-covid-19-insurance-scams#.Xts8DzozbIX. [Accessed: 06-Jun-2020].
7. "Indian general insurance market grows 14.5% | Insurance Business." [Online]. Available: https://www.insurancebusinessmag.com/asia/news/breaking-news/indian-general-insurance-market-grows-14-5-213662.aspx. [Accessed: 20-Sep-2020].
8. S.-H. Li, D. C. Yen, W.-H. Lu, and C. Wang, "Identifying the signs of fraudulent accounts using data mining techniques," *Comput. Human Behav.*, vol. 28, no. 3, pp. 1002–1013, May 2012, doi: 10.1016/j.chb.2012.01.002.
9. N. Pillay, A. P. Engelbrecht, A. Abraham, M. C. Du Plessis, V. Snášel, and A. K. Muda, *Advances in Nature and Biologically Inspired Computing*, vol. 419. Cham: Springer International Publishing, 2016.
10. "How European Insurers Can Manage The Impact Of Covid-19." [Online]. Available: https://www.oliverwyman.com/our-expertise/insights/2020/mar/covid-19-european-insurance.html. [Accessed: 05-Jun-2020].
11. "COVID-19 Impact on Global Insurance Fraud Detection Industry 2020: Industry Trends, Size, Share, Growth Applications, SWOT Analysis by Top Key Players and Forecast Report to 2026 – Jewish Market Reports." [Online]. Available: https://jewishlifenews.com/uncategorized/covid-19-impact-on-global-insurance-fraud-detection-industry-2020-industry-trends-size-share-growth-applications-swot-analysis-by-top-key-players-and-forecast-report-to-2026/. [Accessed: 05-Jun-2020].
12. "Impact of COVID-19 on Life And Health Insurance | TCS." [Online]. Available: https://www.tcs.com/blogs/how-technology-can-aid-insurers-combat-covid-19. [Accessed: 27-May-2020].
13. "COVID-19 Impact on Fraud Detection and Prevention (FDP) Market | Coronavirus Outbreak & FDP Industry | MarketsandMarkets." [Online]. Available: https://www.marketsandmarkets.com/Market-Reports/covid-19-impact-on-fraud-detection-and-prevention-market-23997778.html. [Accessed: 05-Jun-2020].
14. "COVID-19 online fraud trends: Industries, schemes and targets - Help Net Security." [Online]. Available: https://www.helpnetsecurity.com/2020/05/15/covid-19-online-fraud/. [Accessed: 05-Jun-2020].
15. "COVID-19 (C-19) and Fraudulent Claims."
16. "COVID-19 – A backdoor to increased fraud risk?" [Online]. Available: https://www2.deloitte.com/ch/en/pages/financial-advisory/articles/covid-19-operating-in-the-new-normal-fraud-risk.html. [Accessed: 05-Jun-2020].
17. N. Dragano, C. J. Rupprecht, O. Dortmann, M. Scheider, and M. Wahrendorf, "Higher risk of COVID-19 hospitalization for unemployed: an analysis of 1,298,416 health insured individuals in Germany," *medRxiv*, p. 2020.06.17.20133918, Jun. 2020, doi: 10.1101/2020.06.17.20133918.
18. J. W. Goodell, "COVID-19 and finance: Agendas for future research," *Financ. Res. Lett.*, vol. 35, Jul. 2020, doi: 10.1016/j.frl.2020.101512.
19. "Evaluating the impact of COVID-19 on insurance sector technology - Omdia." [Online]. Available: https://technology.informa.com/623267/evaluating-the-impact-of-covid-19-on-insurance-sector-technology. [Accessed: 05-Jun-2020].
20. "How Will COVID-19 Impact Personal Injury Cases?" [Online]. Available: https://www.atltriallaw.com/how-will-covid-19-impact-personal-injury-cases/. [Accessed: 05-Jun-2020].
21. R. Y. Gupta, S. Sai Mudigonda, P. K. Kandala, and P. K. Baruah, "Implementation of a Predictive Model for Fraud Detection in Motor Insurance using Gradient Boosting Method and Validation with Actuarial Models," in *2019 IEEE International Conference on Clean Energy and Energy Efficient Electronics Circuit for Sustainable Development (INCCES)*, 2019, pp. 1–6, doi: 10.1109/INCCES47820.2019.9167733.
22. V. Chandola, "Anomaly Detection : A Survey," 2009.
23. D. Cutler, "How Will COVID-19 Affect the Health Care Economy?," *JAMA - Journal of the American Medical Association*, vol. 323, no. 22. American Medical Association, pp. 2237–2238, 09-Jun-2020, doi: 10.1001/jama.2020.7308.
24. "Insurance sector in India: Overview, IRDAI, Companies, Stats & Trends." [Online]. Available: https://www.acko.com/articles/general-info/insurance-sector-india/. [Accessed: 08-Jun-2020].
25. "Insurance Sector in India: Industry Overview, Market Size & Trends | IBEF." [Online]. Available: https://www.ibef.org/industry/insurance-sector-india.aspx. [Accessed: 08-Jun-2020].
26. "Potential Impact Of COVID-19 On Insurance Fraud Detection Market | Leading Companies Analysis 2027 – Cole Reports." [Online]. Available:

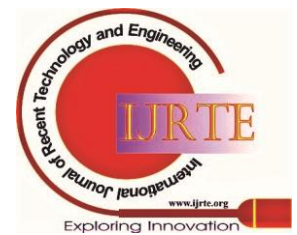

https://coleofduty.com/news/2020/05/23/potential-impact-of-covid-1 9-on-insurance-fraud-detection-market-leading-companies-analysis-2 027/. [Accessed: 05-Jun-2020].

27. [27]   P. Babuna, X. Yang, A. Gyilbag, D. A. Awudi, D. Ngmenbelle, and D. Bian, "The Impact of COVID-19 on the Insurance Industry," *Int. J. Environ. Res. Public Health*, vol. 17, no. 16, p. 5766, Aug. 2020, doi: 10.3390/ijerph17165766.

28. [28]   E. Summary, T. Nadu, I. A. I. Office, and B. T. Results, "COVID-19 A Study and Projections for India - An Update."

29. [29]   R. Y. Gupta, S. S. Mudigonda, P. K. Kandala, and P. K. Baruah, "A Framework for Comprehensive Fraud Management using Actuarial Techniques," vol. 10, no. 3, pp. 780–791, 2019.

30. [30]   N. Rai, P. K. Baruah, S. S. Mudigonda, and P. K. Kandala, "Fraud Detection Supervised Machine Learning Models for an Automobile Insurance," *Int. J. Sci. Eng. Res.*, vol. 9, no. 11, pp. 473–479, 2018.

31. [31]   "Rise in searches for 'How to set fire' a sign insurance fraud beckons as economy crashes." [Online]. Available: https://www.washingtonexaminer.com/news/rise-in-searches-for-how -to-set-fire-a-sign-insurance-fraud-beckons-as-economy-crashes. [Accessed: 06-Jun-2020].

32. [32]   "COVID-19 in India | Kaggle." [Online]. Available: https://www.kaggle.com/sudalairajkumar/covid19-in-india. [Accessed: 19-Sep-2020].

## AUTHORS PROFILE

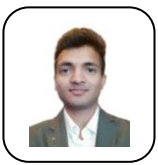

**Rohan Yashraj Gupta,** He is a doctoral research scholar in Sri Sathya Sai Institute of Higher Learning in the field of Actuarial sciences. His area of research includes Data-Driven Fraud Detection and Prevention using Actuarial Techniques and Technology. He is an Actuarial data science expert and has published research papers in international journals.

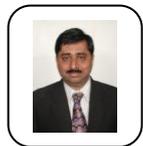

**Satya Sai Mudigonda,** A Senior Tech Actuarial Consultant providing services in USA and India. With a wide skill set, he managed numerous multi-million-dollar international assignments for major insurance companies across the globe. He is an honorary professor in Sri Sathya Sai Institute of Higher Learning. He has published about fifteen papers in the field of Actuarial data science and has presented in several international conferences.

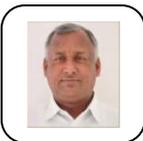

**Dr. Pallav Kumar Baruah,** He is an Associate Professor and the former HOD of the Department of Mathematics and Computer Science of Sri Sathya Sai Institute of Higher Learning. He has guided several research scholars in mathematics and computer science. He has numerous research publications to his credit and has presented in several national and international conferences.

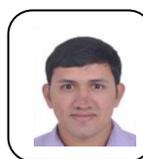

**Phani Krishna Kandala,** an amalgam of Actuarial and Data Science experience in international and multicultural environment. Expertise in Insurance Data Science, Actuarial and Risk Management. Mentors Tech Actuarial Projects for Masters Program in Computer Science. He is a visiting faculty in Sri Sathya Sai Institute of Higher Learning. He has published numerous papers in the field of Actuarial data science and has presented in several international conferences.